\documentclass[prl,aps,twocolumn,showpacs,superscriptaddress]{revtex4}
\usepackage{amsmath,graphicx}
\newcommand{\vect}[1]{\mathbf{#1}}

\begin{document}

\newcommand{\ALPHA}{\alpha}
\newcommand{\SIGMA}{\sigma}

\author{Jonathan Keeling}
\affiliation{Cavendish Laboratory, University of Cambridge,
  J.J.Thomson Ave., Cambridge CB3 0HE, UK}
\author{Natalia G. Berloff}
\affiliation{Department of Applied Mathematics and Theoretical Physics,
  University of Cambridge, Cambridge, CB3 0WA, UK}

\title{Spontaneous rotating vortex lattices in a pumped decaying
    condensate}

\begin{abstract}
  Injection and decay of particles in an inhomogeneous quantum
  condensate can significantly change its behaviour.  We model
  trapped, pumped, decaying condensates by a complex
  Gross-Pitaevskii equation and analyse the density and currents in
  the steady state. With homogeneous pumping, rotationally symmetric
  solutions are unstable.  Stability may be restored by a finite
  pumping spot. However if the pumping spot is larger than the
  Thomas-Fermi cloud radius, then rotationally symmetric solutions are
  replaced by solutions with spontaneous arrays of vortices.  These
  vortex arrays arise without any rotation of the trap, spontaneously
  breaking rotational symmetry.
\end{abstract}
\maketitle

While much of the possible physics of quantum condensates has been
examined in experiments on atomic gases, superfluid Helium and
superconductors, there has recently been much interest in examples of
condensates of quasiparticle excitations, such as
excitons~\cite{butov04,yang06} (bound electron-hole pairs),
exciton-polaritons~\cite{kasprzak06:nature,snoke06:condensation,lagoudakis08}
(superpositions of quantum well excitons and microcavity photons), and
magnons (spin-wave excitations) both in magnetic insulating
crystals~\cite{demokritov06,demidov08} \footnote{Magnon condensation
  experiments have included both cases out of equilibrium with magnon
  injection~\cite{demokritov06,demidov08}, and also those using an
  applied magnetic field as an effective chemical potential (see
  e.g.~\cite{giamarchi08} and Refs.  therein); it is the former of
  these cases which is relevant here.} and in superfluid
$^3$He~\cite{bunkov07,bunkov08,volovik08}.
One particular difference shown by these systems is that the
quasiparticles have finite lifetimes, and as a result, they can be
made to form condensates out of equilibrium, which are best understood
as a steady state balance between pumping and decay, rather than true
thermal equilibrium.
The effects of pumping and decay in these condensates have been the
subject of several recent
works~\cite{tassone99,malpuech02:prb,porras02,doan05:prb,wouters06,wouters06b,
  wouters07:prl,wouters08,szymanska06:prl,szymanska06:long,lagoudakis08}
which have shown that even when collisions can rapidly thermalise the
energy distribution of a system, there may yet be noticeable effects
associated with the energy scale introduced by the pumping and decay.

The Gross-Pitaevskii equation (GPE) has been applied to successfully
describe many features of equilibrium condensates when far in the
condensed regime, including density profiles, the dynamics of
vortices, hydrodynamic modes --- see e.g.~\cite{pitaevskii03} and
Refs.\ therein.
Using a mean-field description of the condensate,
e.g.~\cite{szymanska06:prl,szymanska06:long,wouters08}, one can
recover a complex Gross-Pitaevskii equation (cGPE), including terms
representing gain, loss and an external trapping potential.
This letter studies the interplay between pumping and decay and the
external trapping potential in the context of the cGPE in order to
illustrate some of the differences between equilibrium and
non-equilibrium condensates.
In the absence of trapping, this is the celebrated complex
Ginzburg-Landau equation that describes a vast variety of
phenomena~\cite{aranson02} from nonlinear waves to second-order phase
transitions, from superconductivity to liquid crystals and cosmic
strings and binary fluids~\cite{PhysRevLett.83.3422}.
What is of interest in this letter is how pumping and decay, described
in the cGPE modify behaviour compared to the regular GPE as is widely
applied to spatially inhomogeneous equilibrium quantum
condensates~\cite{pitaevskii03}.
Spatial inhomogeneity, due to either engineered and disorder
potentials, has been studied for both excitons~\cite{butov04} and
polaritons~\cite{richard05:prb,kasprzak06:nature,snoke06:condensation,lagoudakis08}.

By looking for steady state solutions to the cGPE, we find that a
density-dependent gain rate combined with spatial inhomogeneity leads
to steady-state currents, connecting regions of net gain with those
of net loss.
These supercurrents in turn affect the density profile (as is already
well known in the case of solutions with vorticity), and so pumping
and decay can significantly alter the density profile of a trapped
condensate.
The effects of steady-state current flows in the absence of pumping were
considered in Refs.~\cite{porras04,alexandrescu06}.
By studying the stability of these steady state solutions, one finds
that with homogeneous pumping these solutions become unstable to
breaking of rotational symmetry.
Stability can be restored by reducing the size of the pumping spot to
be comparable to the self-consistent size of the condensate cloud (set
by the balance of pumping and decay).
By increasing the pump spot size (or by decreasing the pump strength),
the rotationally symmetric solutions again become unstable, and are
replaced by solutions with vortex lattices.
The observation of vortices driven by the combination of particle flux
and spatial inhomogeneity has been seen
experimentally~\cite{lagoudakis08}; our results indicate that such
vortex solutions can arise even with symmetric traps.
Our findings show the existence of new phenomena in the already rich
world of complex Ginzburg-Landau equations~\cite{aranson02} that play
an enormous role in our understanding of non-equilibrium physics and
pattern formation~\cite{cross93}.

Our cGPE can be derived as the gradient expansion of the saddle point
equation of a non-equilibrium path-integral theory of polariton
condensation~\cite{szymanska06:prl,szymanska06:long}.
However, to provide insight into its meaning, we instead describe
here the physical origin of the terms it contains.
The form of the cGPE depends on whether one considers
coherent or incoherent pumping.
Coherent pumping, injecting particles directly into the condensate at
an energy $\omega_0$, is described by a source term $\partial_t \psi =
F e^{i\omega_0 t}$ ~\cite{wouters07:prb}.
We instead consider non-resonant pumping, and thus we introduce
stimulated scattering into the condensate, $\partial_t \psi
|_{\text{gain}} = \gamma \psi$.
A similar term $\partial_t \psi |_{\text{loss}} = -\kappa \psi$
describes particle decay, i.e. loss, and so we introduce
$\gamma_{\text{eff}} = \gamma - \kappa$.
With such gain and loss, the dynamics is unstable and trivial; if gain
exceeds loss, the condensate grows indefinitely, if loss exceeds gain,
the condensate vanishes.
In practice, for non-resonantly pumped solid-state systems, the gain
is saturable --- it tries to bring the condensate density into
chemical equilibrium with some external particle density.
The simplest model of such a process is a density-dependent rate of
gain, $\partial_t \rho |_{\text{gain}} = (\gamma - \Gamma \rho) \rho$,
which tries to establish equilibrium at $\rho = \gamma/\Gamma$.
A closely related model of saturation, considering a reservoir of
non-condensed particles was studied in~\cite{wouters07:prl}; the steady
state behaviours of both models are very similar.
We combine these terms and write the complex GPE in the following form
\begin{equation}
  \label{eq:GPE}
  i\hbar\partial_t \psi
  =
  \left[
    - \frac{\hbar^2 \nabla^2}{2m} + V(r) 
    + U |\psi|^2
    + i (\gamma_{\text{eff}} - \Gamma |\psi|^2 )
  \right]
  \psi,
\end{equation}
where $V(r)$ is an external trapping potential, and $U$ is the
strength of the $\delta$-function interaction (pseudo) potential.

We will we look for steady state solutions and introduce the chemical
potential, $\mu$, in the usual way, via $i\hbar
\partial_t \psi(t) = \mu \psi(t)$.
In this equation $\mu$ is a free parameter to be determined from the
balance of gain and loss; neither the chemical potential nor total
number of particles is externally imposed.
We will illustrate how the interaction of spatial inhomogeneity with
pumping and decay modifies the density profile by studying how the
profile depends on pumping strength in a number of cases.

We consider the classic example of a parabolic trapping
potential in two dimensions.
For this problem, two dimensionless parameters control the behaviour.
We can write the potential as $V(r) = (\hbar \omega/2) (r^2/l^2)$,
where $\omega$ is the oscillator frequency and $l =
\sqrt{\hbar/m\omega}$ is the oscillator length.
Expressing lengths in units of $l$, energies in units of $\hbar
\omega$, and rescaling $\psi \to \sqrt{\hbar \omega/2 U} \psi$,
yields:
\begin{equation}
  \label{eq:GPE_rescale}
  \left( \frac{2\mu}{\hbar \omega} \right)
  \psi
  =
  \left[
    - \nabla^2 + r^2
    + |\psi|^2
    + i \left(\frac{2\gamma_{\text{eff}}}{\hbar \omega} - \frac{\Gamma}{U} |\psi|^2 \right)
  \right]
  \psi.
\end{equation}
For the rest of this letter, we shall write $\tilde{\mu} = 2 \mu/\hbar
\omega$, and introduce the two dimensionless parameters which control
the density profile: $\ALPHA =2\gamma_{\text{eff}}/\hbar \omega$, and
$\SIGMA = \Gamma/U$.

Before discussing the solutions, we give illustrative values
$\alpha,\sigma$, relevant to the polariton experiments of
Refs.~\cite{richard05:prb,kasprzak06:nature}.
The maximum pumping strengths considered are around ten times the
threshold pumping strength; this threshold occurs when pumping matches
decay rate $\gamma=\kappa$.
The decay rate, found from the linewidth at low power, is $\kappa
\simeq 0.13$meV, and so the pump rate may be up to
$\gamma_{\text{eff}} < 1.2$meV.
To find $\alpha$, one needs also the characteristic trap scale.
In Ref.~\cite{richard05:prb}, the disorder traps are estimated to have
a depth $E_0 \simeq 0.5$meV and size $a \simeq 3\mu$m, which with a
polariton mass of $m \simeq 10^{-4} m_0$ yields a trap frequency
$\hbar \omega = \sqrt{E_0 \hbar^2/m a^2} \simeq 0.2$meV; hence $0 \le \alpha \lesssim 10$.
$\Gamma$ is harder to estimate without a specific microscopic model;
an order-of-magnitude estimate may be found from the observed blue
shift (shift of chemical potential) vs pumping power.
As discussed below, for weak pumping one has $\mu \simeq (\hbar
\omega/2) (3 \alpha / 2 \sigma)$, and so $\sigma \simeq 3
\gamma_{\text{eff}} / 2 \mu$.
In Ref.~\cite{kasprzak06:nature}, a pump power at twice threshold,
i.e.  $\gamma_{\text{eff}} \sim 0.13$meV yields a blue shift $\mu\sim
0.5$meV, giving $\sigma \simeq 0.3$, however this estimate involves
considerable uncertainty.

\begin{figure}[htpb]
\centering
\includegraphics[width=0.90\columnwidth]{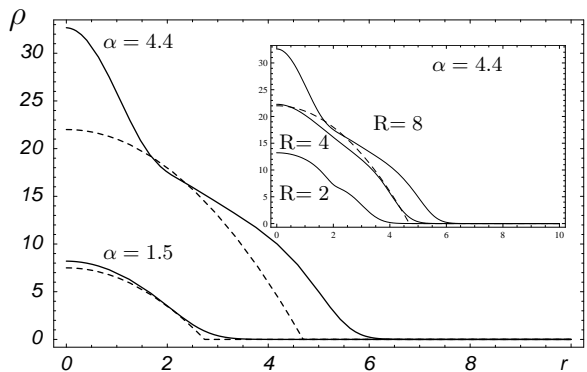}
\caption{The densities, $\rho(r)$, of the steady states of 
  Eq.~(\ref{eq:GPE_rescale}) for $\sigma=0.3$ and $\alpha=1.5, 4.4$
  (black solid lines) as compared to the TF solutions $\rho=\mu^*-r^2,
  r < \sqrt{\mu^*}$ (dashed lines). The inset shows the symmetric
  solutions for $\alpha=4.4$ with finite pump spot size as labelled;
  $R=2,4$ are stable, while $ R=8$ is unstable to breaking rotational
  symmetry.  }
\label{profile}
\end{figure}

We first discuss the rotationally symmetric steady states of
Eq.~(\ref{eq:GPE_rescale}), using fixed point iterations combined with
the secant algorithm for determining $\tilde{\mu}$ for a variety of
parameters.
We compare the densities of the ground state with the analytical
Thomas-Fermi (TF) profiles found by neglecting density gradients and
assuming that supercurrents do not affect the density distributions.
Figure~\ref{profile} shows the density profiles for different values of
$\ALPHA$. 
As $\ALPHA$ increases, two effects are clear: firstly the increased
pumping rate evidently leads to an increased total density of the
condensate; secondly increased pumping leads to a greater flux, and so
for $\alpha=4.4$, the density profile is not the TF profile, but is
suppressed in the middle, where the supercurrent is highest.
The increase in total density can be described from the balance of net
gain and loss; by multiplying Eq.~(\ref{eq:GPE_rescale}) by
$\psi^{\ast}$ and integrating over all space, the imaginary part of
this equation is:
\begin{equation}
  \label{eq:identity}
  \int d^2 r
  \left(
    \ALPHA - \SIGMA |\psi|^2
  \right) |\psi|^2
  =
  0.
\end{equation}
When pumping is not too strong, substituting the Thomas-Fermi solution
$|\psi|^2 = (\tilde{\mu} - r^2)$ for $r < \sqrt{\tilde{\mu}}$ into this equation
yields $\tilde{\mu} = \mu^*\equiv 3 \alpha/2 \sigma$.

The suppression of density due to supercurrent means that with
increasing pumping, the density profile becomes increasingly sharp as
supercurrent flows become important.
Such results have been strongly hinted at in several microcavity
polariton experiments where sharpening of the peaks of the density
profile with increasing density is seen with both disorder
traps~\cite{richard05:prb,kasprzak06:thesis} and engineered stress
traps~\cite{snoke06:condensation}.

Let us now discuss how the changes to the density profile
seen above can be understood physically.
The Madelung transformation, $\psi = \sqrt{\rho} e^{i\phi}$,
represents Eq.~(\ref{eq:GPE_rescale}) as a continuity equation and
Bernoulli's equation:
\begin{align}
  \label{eq:rho-v-ip}
  \nabla\cdot[\rho\nabla\phi] 
  =
  \left(
    \ALPHA - \SIGMA \rho 
  \right) \rho,
  \\
  \label{eq:rho-v-rp}
  \tilde{\mu}
  = 
  \left|\nabla \phi \right|^2 
  + r^2 + \rho - \frac{\nabla^2 \sqrt{\rho}}{\sqrt{\rho}}.
\end{align}
Regions of high density imply loss, and regions of low density gain,
which lead to supercurrents $\nabla \phi$, between these regions.
If these supercurrents are large, they affect the Bernoulli equation,
leading to a density depletion where current is largest.
This is clearly seen in Fig.~\ref{profile}; there is net gain at large
radii, and net loss at small radii, and a dip in the density profile
in between these indicates a region of maximum supercurrent.
The radial phase gradient associated with the current could be seen
experimentally using interferograms as in
e.g.~\cite{kasprzak06:nature,kasprzak06:thesis,lagoudakis08}; the
maximum phase difference across the cloud scales as $\Delta \phi
\propto \sigma \mu^2$ and $\Delta \phi \simeq 30$, for relevant
parameters.

To study stability, the time evolution of Eq.~(\ref{eq:GPE}) is
followed, using the rotationally symmetric steady-state solutions as
initial conditions, and including a small perturbation.
With an infinite homogeneous pump, as in Eq.~(\ref{eq:GPE}), the
solution is always unstable to angular perturbations.  
This instability can also be seen by considering pumping and decay as
perturbative corrections to the hydrodynamic modes of a trapped 2D
condensate; one finds the leading order correction to the mode
energies introduces growth/decay rates which always produce growth for
modes with large enough angular momentum.
Physically, this instability can be understood by looking at the
region just outside the condensate cloud.
In this region the steady state gain is zero since it is proportional
to density, however linear stability analysis for $\rho \to \rho +
\delta \rho$ depends on $\partial_{\rho} \left[ (\alpha - \sigma \rho)
  \rho \right] \propto \alpha - 2 \sigma \rho$, which is positive
outside the condensate cloud, so any small perturbation will grow.
High angular momentum hydrodynamic modes of the condensate are
unstable because they transfer density to the edge of the condensate.

\begin{figure}[htpb]
  \centering
  \includegraphics[width=0.95\columnwidth]{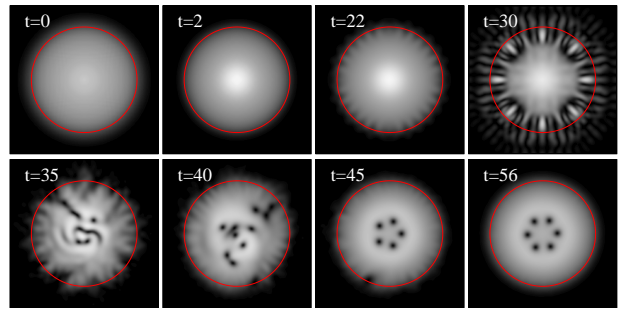}
  \caption{(Color online) Time evolution from the rotationally
    symmetric steady state solution of Eq.~(\ref{eq:GPE_rescale}) for
    $\sigma=0.3, \alpha=4.4$ when the radius of the finite pumping
    spot is $R=5$ (as marked by the red line).  Times are in units of
    $2/\omega$, were $\omega$ is frequency of harmonic trap.}
  \label{fig:storybook}
\end{figure}

This mechanism of instability is supported by observing that the
instability is not present with a finite spot size; this ensures that
outside the condensate cloud there is no gain, and so no growth.
For simplicity, we treat this radial cutoff by replacing $\alpha$ by
$\alpha(r) = \alpha \Theta(R- r)$, where $\Theta$ is the unit step function
and $R$ the cutoff radius.
A finite spot, of size comparable to the observed cloud is in fact
used in current
experiments~\cite{kasprzak06:nature,snoke06:condensation,lagoudakis08}.
For small $R$, this stabilises the radially symmetric modes.
However, when $R$ exceeds the Thomas-Fermi condensate radius,
$\sqrt{\tilde{\mu}} \simeq \sqrt{ 3 \alpha / 2 \sigma}$, the
instability reappears.
The subsequent time dynamics, leading to a new steady state is shown
in Fig.~\ref{fig:storybook}.
The final state is no longer stationary, but instead 
rotates according to: $i \hbar
\partial_t \psi = (\mu - 2\Omega L_z )\psi$, where $L_z = i (x
\partial_y - y \partial_x)$.

\begin{figure}[htpb]
  \centering
  \includegraphics[width=0.94\columnwidth]{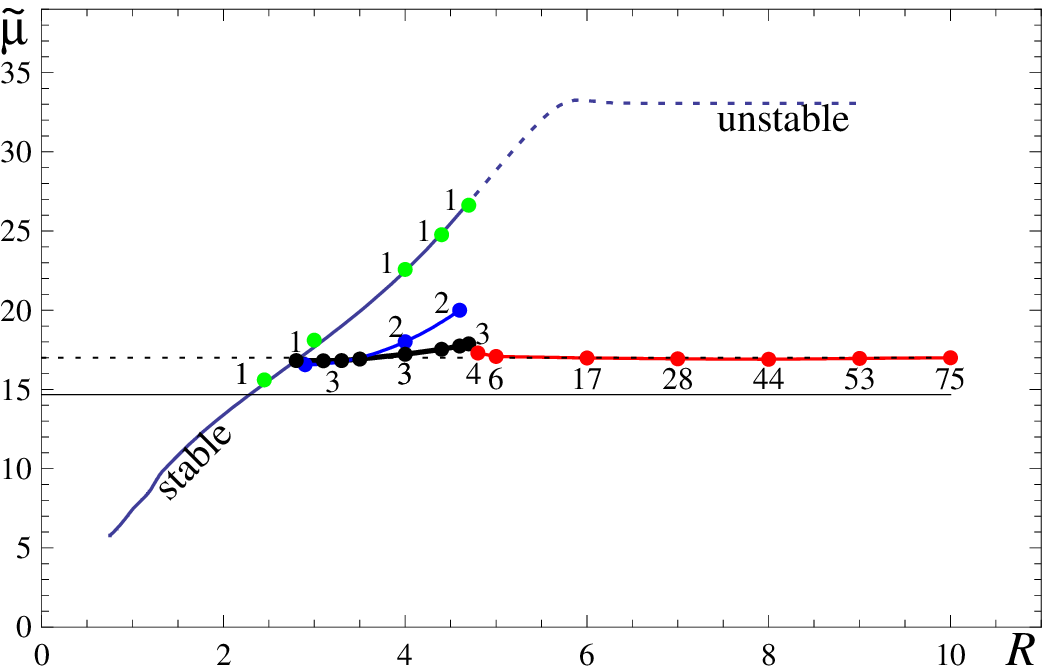}%
  \raisebox{0.34\columnwidth}{
    \hspace{-0.90\columnwidth}
    \includegraphics[width=0.25\columnwidth]{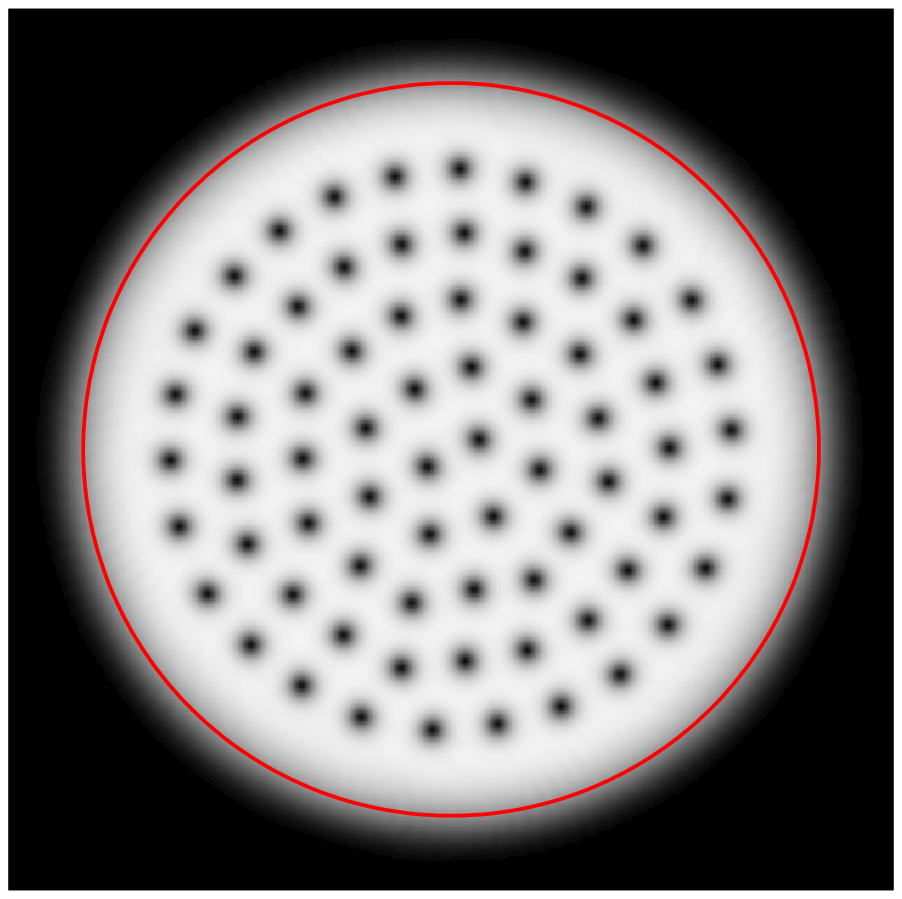}%
    \hspace{0.65\columnwidth}}
  \caption{(Color online) Chemical potential $\tilde \mu$ as a
    function of pumping spot size $R$. Numbers of vortices for the
    stable vortex lattice configuration are marked. Solid lines join
    branches of stable solutions. Dotted line corresponds to unstable
    radially symmetrical solutions without vortices.  Horizontal solid
    line marks $\tilde{\mu} = \alpha/\sigma$.  Inset: vortex lattice
    for $R=10$.}
  \label{fig:muR}
\end{figure}

As the initial problem is rotationally symmetric and non-rotating, the
vortex solution spontaneously breaks rotation symmetry; either sign of
vortex array is stable, but the rotationally symmetric solution is not
stable.
This behaviour is characteristically different from the equilibrium
non-rotating trapped condensate in which vortex solutions are
unstable~\cite{PhysRevLett.79.2164}, and vortices would spiral out of
the condensate --- with pumping and decay the dynamics shown in
Fig.~\ref{fig:storybook} shows that vortices spiral into the condensate.
In addition, for a given radius of pump spot, more than one vortex
array may be stable, the number of vortices depending on the history
of the spot size; this is indicated in Fig.~\ref{fig:muR}. This is
similar to hysteresis effects in rotating Bose-Einstein
condensates~\cite{barenghi06}, but the external rotation is absent in
the model considered.
The origin of the instability --- growth of condensate density outside
the Thomas-Fermi radius --- suggests that other models of the cGPE
with a reservoir would show the same behaviour~\cite{wouters08}; for
the instability to be removed, one requires a reservoir concentrated
near the minimum of the trap.
In the context of the polariton condensate, this means the instability
might be cured if the non-condensed exciton reservoir was highly
mobile --- in the language of laser theory, this corresponds to
damping of instabilities by carrier diffusion.

This solution of the cGPE can be understood as vortices enlarging the
cloud size to match the pump spot.
Adapting Eqs.~(\ref{eq:rho-v-ip},\ref{eq:rho-v-rp}) for a rotating
solution~\cite{pitaevskii03} gives:
\begin{align}
  \label{eq:rho-v-rot-ip}
  \nabla\cdot[\rho(\nabla\phi - \Omega \times \vect{r})] 
  =
  \left(
    \ALPHA\Theta(R-r) - \SIGMA \rho 
  \right) \rho,
  \\
  \label{eq:rho-v-rot-rp}
  \tilde{\mu}
  \simeq
  \left|\nabla \phi - \Omega \times \vect{r} \right|^2 
  + r^2(1-\Omega^2) + \rho - \frac{\nabla^2 \sqrt{\rho}}{\sqrt{\rho}}.
\end{align}
The rotating vortex lattice solution adopted can be understood as
follows; vortices lead to quantised rotation, and the density of
vortices, $n_{\text{v}} \simeq \Omega/\pi$ ensures that $\nabla \phi
\simeq \Omega \times \vect{r}$ mimicking solid body rotation.
Neglecting the vortex core, the continuity equation,
Eq.~(\ref{eq:rho-v-rot-ip}), thus requires $\rho \simeq \alpha/\sigma$.
This implies that inside the vortex lattice there is no net radial
current in contrast to the solutions with smaller R; for the solutions
with a single vortex (when $R<\sqrt{3 \alpha/2 \sigma}$), the
combination of radial and rotational currents means such vortices
are in fact ``spiral vortices''.
For this constant density solution to be valid (except near each
vortex core) Eq.~(\ref{eq:rho-v-rot-rp}) requires $\Omega \simeq 1$
and $\tilde{\mu} = \rho \simeq \alpha/\sigma$.
This solution persists till the edge of the vortex lattice, beyond
which $\nabla \phi = N_{\text{v}}/r$, where $N_\text{v}$ is the total number
of vortices.
The total number of vortices is then set by requiring the edge of the
cloud to occur around $r=R$, leading to $N_\text{v} \simeq
n_{\text{v}} \pi R^2 \simeq R^2$ for large $R$.
When the vortex core is not negligible, the extra gain in the vortex
core and quantum pressure corrections imply $\mu > \rho >
\alpha/\sigma$.
This is shown quantitatively in Fig.~\ref{fig:muR}.

For yet larger $R$,
e.g. $R=20$, no such simple rotating vortex lattice is found --- for
such parameters there is a residual vortex lattice in the center of
the cloud, but the behaviour at the edge becomes irregular.

In conclusion, we have shown that steady-state currents connecting
regions of net gain and loss can lead to significant modifications of
the density profile of a quantum condensate, leading even to instability
of the rotationally symmetric state and the spontaneous creation of a 
vortex array.
Vortices can be clearly observed as in Ref.~\cite{lagoudakis08} by
pairs of forks in the interferogram of the emitted light.
To observe the spontaneous vortex array, one would require that
disorder is weak compared to the harmonic trap, which may prevent its
observation in the current generation of semiconductor microcavities,
however other than this hurdle, the numerical estimates place current
polariton experiments in a regime in which such effects could occur.

\acknowledgments{J.K.\ acknowledges discussions with N.~R.~Cooper,
  I.~Carusotto, P.~B.~Littlewood and M.~H.~Szyma\'nska, and financial
  support from Pembroke College, Cambridge. N.G.B.\ acknowledges
  discussion with Erich Mueller and financial support from the
  EPSRC-UK.  }


\end{document}